\documentclass[reprint,
 amsmath,amssymb,
 aps,
]{revtex4-2}

\usepackage{graphicx, color}
\usepackage{dcolumn}
\usepackage{bm}
\usepackage[mathlines]{lineno}
\usepackage{amsmath}
\usepackage{amssymb}
\usepackage{mathrsfs}
\usepackage{float}
\usepackage{soul}
\usepackage{lipsum}
\usepackage{hyperref}
\usepackage[normalem]{ulem}

\hypersetup{
    colorlinks=true,       
    linkcolor=green2,          
    citecolor=green2,        
    filecolor=magenta,      
    urlcolor=green2           
}
\definecolor{rp}{cmyk}{0.2, 1, 0.6, 0}
\definecolor{green2}{cmyk}{0.27, 0, 1, 0.52}
\usepackage{cleveref}

\begin{document}

\preprint{KCL-PH-TH/2024-08}

\title{Engineering and Revealing Dirac Strings in Spinor Condensates} 

\author{Gui-Sheng Xu$^{\text{1,2,3}}$}
\author{Mudit Jain$^{\text{4,5}}$}%
\email{mudit.jain@kcl.ac.uk}
\thanks{First Author and Second Author contributed equally to this work.}
\author{Xiang-Fa Zhou$^{\text{1,2,3}}$}%
\email{xfzhou@ustc.edu.cn}
\author{Guang-Can Guo$^{\text{1,2,3}}$}
\author{Mustafa A. Amin$^{\text{4}}$}
\author{Han Pu$^{\text{4}}$}%
\email{hpu@rice.edu}
\author{Zheng-Wei Zhou$^{\text{1,2,3}}$}%
 \email{zwzhou@ustc.edu.cn}
\address{$^{\text{1}}$CAS Key Lab of Quantum Information, University of Science and Technology of China, Hefei, 230026, China\\
$^{\text{2}}$Synergetic Innovation Center of Quantum Information and Quantum Physics, University of Science and Technology of China,
Hefei, 230026, China\\
$^{\text{3}}$Hefei National Laboratory, Hefei 230088, China\\
$^{\text{4}}$Department of Physics and Astronomy, Rice University, Houston, Texas 77251, USA\\
$^{\text{5}}$Theoretical Particle Physics and Cosmology, King's College London, Strand, London, WC2R 2LS, United Kingdom
}

\date{\today}

\begin{abstract}
Artificial monopoles have been engineered in various systems, yet there has been no systematic study of the singular vector potentials associated with the monopole field. We show that the Dirac string, the line singularity of the vector potential, can be engineered, manipulated, and made manifest in a spinor atomic condensate. We elucidate the connection among spin, orbital degrees of freedom, and the artificial gauge, and show that there exists a mapping between the vortex filament and the Dirac string. We also devise a proposal where preparing initial spin states with relevant symmetries can result in different vortex patterns, revealing an underlying correspondence between the internal spin states and the spherical vortex structures. Such a mapping also leads to a new way of constructing spherical Landau levels, and monopole harmonics. Our observation provides insights into the behavior of quantum matter possessing internal symmetries in curved spaces.
\end{abstract}


\begin{flushleft}
\raggedright
    \small KCL-PH-TH/2024-08
\end{flushleft}

\maketitle


\paragraph*{Introduction ---} Despite of the lack of unambiguous experimental evidence for their existence, magnetic monopoles have a central place in our understanding of quantum matter and modern cosmology~\cite{dirac_QuantisedSingularitiesElectromagnetic_1931,wu_Conceptnonintegrablephase_1975,polyakov_ParticleSpectrumQuantum_1974,hooft_Magneticmonopolesunified_1974}. Remarkably, recent theories and experiments have provided ample evidences for the emergence of artificial monopoles in various physical systems~\cite{bramwell_SpinIceState_2001,ruostekoski_MonopoleCoreInstability_2003,kiffner_MagneticMonopolesSynthetic_2013,sugawa_SecondChernnumber_2018,pietila_CreationDiracMonopoles_2009,savage_Diracmonopolesdipoles_2003,pietila_CreationDiracMonopoles_2009,pietila_NonAbelianMagneticMonopole_2009,ray_ObservationDiracmonopoles_2014a,ray_Observationisolatedmonopoles_2015,stoof_MonopolesAntiferromagneticBoseEinstein_2001,martikainen_CreationMonopoleSpinor_2002,ollikainen_ExperimentalRealizationDirac_2017}. It is well known that the vector potential associated with the monopole field contains line singularities (known as Dirac strings) that terminate at the monopole, even though the monopole magnetic field itself is smooth everywhere (except at the position of the monopole)~\cite{shnir_Magneticmonopoles_2005}. This does not pose any problem as the vector potential, unlike the field, is not ``real" in the sense that it cannot be directly measured. This is also reflected in the fact that the positions of these Dirac strings are gauge dependent.

This conventional wisdom, however, is not necessarily true in systems with artificial gauge fields. In such systems, one often directly realizes and controls the artificial gauge potential, rather than the associated `magnetic' field, rendering the former directly measurable. Indeed, the physical effects of artificial gauge potentials on time-of-flight images of cold atoms have been reported in several experiments ~\cite{dalibard_Artificialgaugepotentials_2011,higbie_PeriodicallyDressedBoseEinstein_2002,papoff_Transientvelocityselectivecoherent_1992,lin_BoseEinsteinCondensateUniform_2009,zhu_SpinHallEffects_2006}.

\begin{figure}
	\centering
 \includegraphics[width=0.33\textwidth]{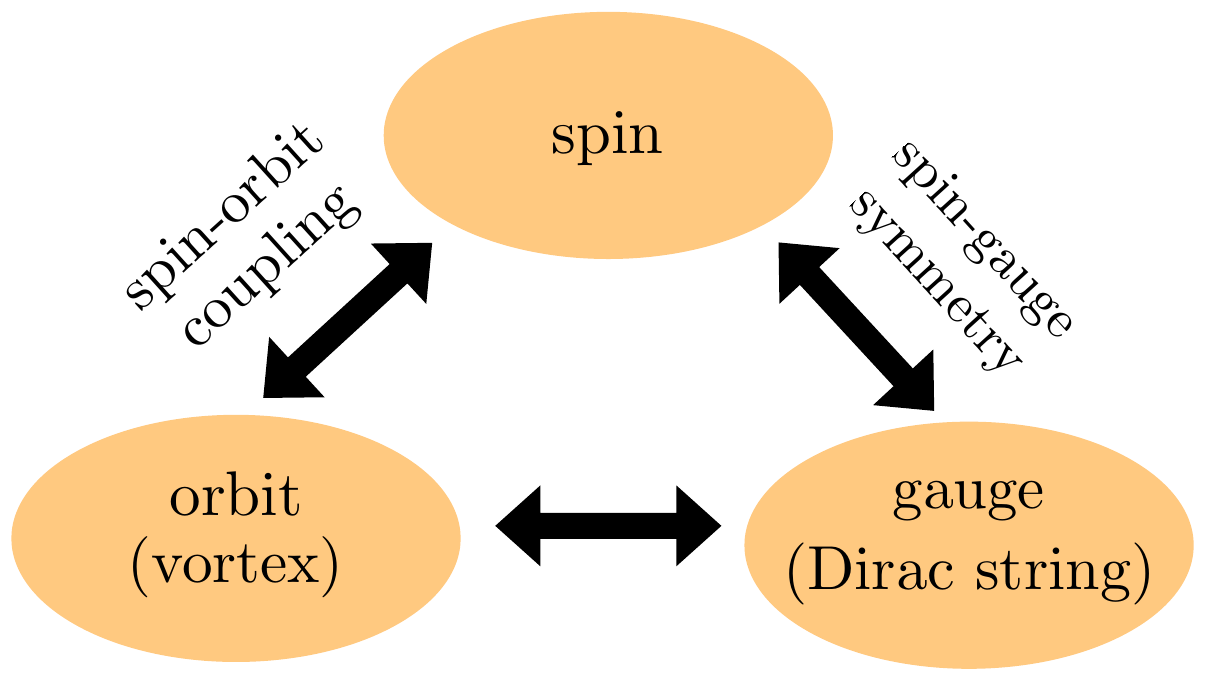}
	\caption{Conceptual plot showing the connection among spin, orbit degrees of freedom, and the artificial gauge field.}
	\label{scheme}
\end{figure}

One widely used platform to realize artificial gauge field is spinor Bose-Einstein Condensates (BECs). When the atomic spin~\footnote{Throughout this work, we shall refer to the hyperfine as spin.} adiabatically follows an external magnetic field, the system accumulates a geometric phase ~\cite{berry_Quantalphasefactors_1984}, which induces an artificial gauge potential governing the spatial wave function; a manifestation of the spin-orbit coupling. If the atoms occupy a spatial region that contains a degenerate point with vanishing magnetic field, the artificial gauge potential can develop a line singularity, which can be regarded as an analog of the Dirac string and a consequence of the local spin-gauge symmetry~\cite{pietila_CreationDiracMonopoles_2009,ho_LocalSpinGaugeSymmetry_1996}. The purpose of this work is to elucidate the relationship among spin, orbit and the artificial gauge field. As conceptually represented in Fig.~\ref{scheme}, through the spin-orbit coupling and the spin-gauge symmetry, there exists a mapping between the vortex filament and the Dirac string.
This directly leads to a novel adiabatic scheme for preparing vortex configurations on a sphere where some initial spin state is prepared for the spinor condensate, and then the artificial magnetic field strength is turned on adiabatically. As a result of conservation of the total angular momentum, some of the initial spin angular momentum is transferred to the orbital degrees of freedom, resulting in the formation of vortices. Preparing different initial spin states can therefore result in different vortex patterns. From the point of view of the artificial gauge field, this amounts to different gauge choices that lead to different Dirac strings.\\

\paragraph*{Spin, vortex, and Dirac strings ---}
\label{II}

Let us consider a spin-$F$ atom of mass $M$ confined in an isotropic harmonic trapping potential with frequency $\omega$, subjected to a hedgehog magnetic field ${\bm B}({\bm r}) \propto {\bm r}$. The realization of the hedgehog field has been proposed in our earlier work~\cite{zhou_SyntheticLandauLevels_2018}. Working in units where $\hbar=M=\omega=1$, the single-particle Hamiltonian of the system reads
\begin{equation}
\label{ham}
    {\rm H}_0(\bm{r})=-\frac{\nabla^2}{2}+\frac{r^2}{2}-\alpha r \hat{\bm{r}}\cdot \mathbf{F}\,.
\end{equation}
Here $\alpha$ characterizes the strength of the hedgehog field, which we assume can be dialled from zero to large values. In the limit of large field strength $\alpha r$, the lowest spin state follows the local magnetic field and satisfies $(\hat{\bm r}\cdot {\bf F})|F_{\hat{\bm r}} \rangle = F|F_{\hat{\bm r}} \rangle$, where $|F_{\hat{\bm r}}\rangle$ can be obtained from $|F_{\hat{\bm z}} \rangle$ (the spin state polarized along the $z$-axis) via rotations: $\left|F_{\hat{\bm r}} \right\rangle=e^{-i \varphi {\rm F}_{\hat{\bm z}}} e^{-i \theta {\rm F}_{\hat{\bm y}}} \left|F_{\hat{\bm z}} \right\rangle$.
Here $\theta$ and $\varphi$ are the polar and azimuthal angles respectively. The total wave function, $\Psi({\bm r})$, can be written as $\Psi({\bm r}) = \psi({\bm r}) |F_{\hat{\bm r}} \rangle$, with the effective Hamiltonian of the scalar wave function $\psi({\bm r})$ being:
\begin{equation}
    {\rm H}_{\rm eff} + \frac{\alpha^2F^2}{2} = \frac{1}{2}(-i\nabla + {\bm A})^2 + \frac{1}{2}(r-\alpha F)^2 + \frac{F}{2r^2}\,.
    \label{heff}
\end{equation}
Here the effective vector potential ${\bm A}=i\langle F_{\hat{\bm r}}|\nabla F_{\hat{\bm r}} \rangle =\hat{e}_\varphi\, F\cos\theta/(r \sin\theta)$ reflects a `magnetic' monopole (of magnetic charge $F$) at the origin. Furthermore, the last term in Eq.~\eqref{heff} dictates that this monopole is also `electrically polarized', with an electric dipole moment $F/2$. The vector potential ${\bm A}$ is singular for $\theta=0$ and $\pi$, which corresponds to two antipodal Dirac strings. It can be seen that the effective system realises a Haldane's sphere~\cite{haldane_FractionalQuantizationHall_1983}: atoms of unit `electric charge', are confined within a thin spherical shell (of order unity width in units of $\sim \sqrt{\hbar/M\omega}$) centered at $r_0 \approx \alpha F$ (in units of $\sim \sqrt{\hbar/M\omega}$), in the presence of an `electrically polarized magnetic monopole' at the origin.

If we rewrite the scalar function above as $\psi({\bm r})=\sqrt{n({\bm r})}e^{i \phi({\bf r})}$, where $n({\bm r})$ represents the local atomic number density, then the total wavefunction can be written as $\Psi({\bm r})=\sqrt{n({\bm r})}|\tilde{F}_{\hat{\bm r}} \rangle$, where $|\tilde{F}_{\hat{\bm r}} \rangle = e^{i \phi({\bf r})}|{F}_{\hat{\bm r}} \rangle$. In other words, we have absorbed the phase factor of the scalar function into a redefinition of the radial spin state. The corresponding vector potential associated with $|\tilde{F}_{\hat{\bm r}} \rangle$ is given by $\tilde{\bm A}=i\langle \tilde{F}_{\hat{\bm r}}|\nabla \tilde{F}_{\hat{\bm r}} \rangle ={\bm A} + \nabla \phi$, which is related to ${\bm A}$ by a gauge transformation. This is just a manifestation of the spin-gauge symmetry in spinor gases~\cite{ho_LocalSpinGaugeSymmetry_1996}. On the other hand the velocity field associated with the total wavefunction $\Psi({\bm r})$ is given by $\boldsymbol{v}^{\text {mass}}=-\nabla \phi - {\bm A} = -\tilde{\bm A}$, which allows us to clearly see the connection between vortices (line singularities of $\boldsymbol{v}^{{\rm mass}}$) and line singularities of $\tilde{\bm A}$. As we will show in the following, by preparing different initial spin configurations in the absence of the monopole field followed by its adiabatic turn on, we may result in final states with different phase structure $\phi({\bf r})$, and hence different vortex or Dirac string orientations. In a sense, changing the initial spin configuration amounts to choosing a different gauge for the vector potential $\tilde{\bm A}$.\\

\begin{figure}
\centering
\includegraphics[width=8.0cm]{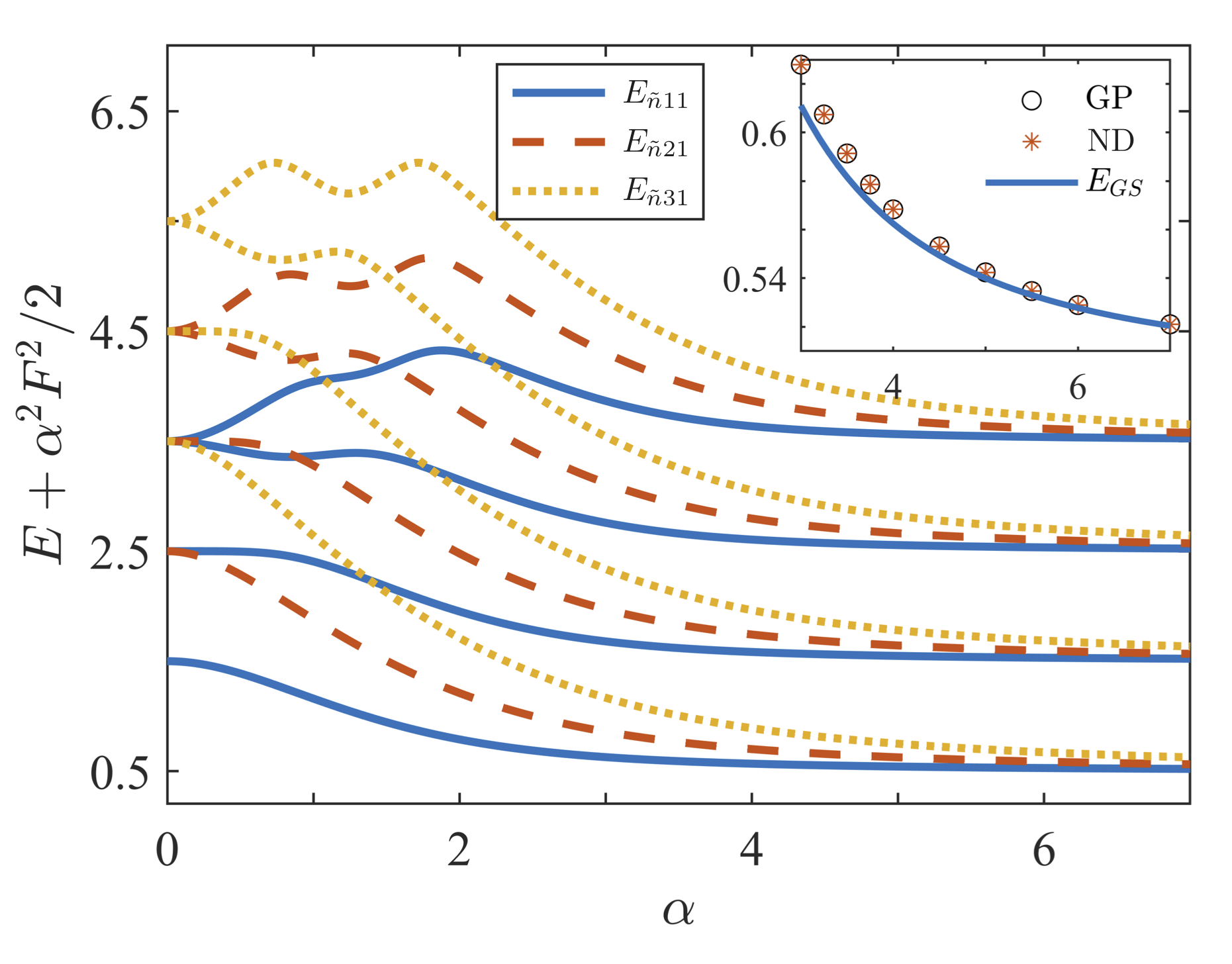}
\caption{The single-particle energy spectrum $E_{\tilde{n}jF}$ (shifted by $\alpha^2 F^2/2$) as a function of $\alpha$, for the $F=1$ case. Upon increasing $\alpha$, different oscillator levels approach each other, and a crossover from the 3D isotropic oscillator levels to Landau levels is seen. The inset shows the lowest energy $E_{011}$ in the adiabatic regime, which is obtained from GP equations (circle), numerical diagonalization (star) and the analysis in Eq.~\eqref{eq:adiabatic_energy} (solid).}
\label{fig:GS_spectrum}
\end{figure}

\paragraph*{Engineering Dirac Strings ---}
To substantiate the argument we laid out above, here we provide a more quantitative description.

\subparagraph*{Single-particle spectrum}: Let us first consider the single-particle spectrum for the Hamiltonian~\eqref{ham}. It can be seen that while for $\alpha = 0$ both ${\bf L}$ and ${\bf F}$ are conserved, for $\alpha \neq 0$ only the total angular momentum (TAM) ${\bf J} = {\bf L} + {\bf F}$ is conserved. This means that the energy eigenstates are also eigenstates of $\{{\bf J}^2, {\rm J}_z\}$. These are the well-known spinor harmonics $|\chi_{j m}^{\ell F}\rangle = \sum^{F}_{m_F=-F} c_{jm}^{\ell m_F} Y^{\ell}_{m-m_F}(\hat{\bm r})|{m_F}_{\hat{\bm z}}\rangle$~\cite{sakurai_ModernQuantumMechanics_2017}, where, $c$'s are the Clebsch-Gordan coefficients, $Y$'s are the usual  spherical harmonics, and $|{m_F}_{\hat{\bm z}}\rangle$ are the spin multiplicity states (in the $z$ basis). Furthermore, while ${\bf F}^2$ is conserved, ${\bf L}^2$ is not. Summing over the $\ell$ quantum number then (and with $\tilde{n}$ as the radial quantum number), the eigenstates take the form
\begin{align}
\label{eq:Psi_full_decompose}
 \Psi_{\tilde{n}jmF}({\bm r})=\sum^{j+F}_{\ell=|j-F|} f_{\tilde{n}jF}^{\ell}(r) |\chi_{j m}^{\ell F} (\hat{\bm r})\rangle\,.
\end{align}
With ($j,m$) being good quantum numbers, and since $\alpha = 0$ corresponds to a simple $3$D harmonic oscillator (with an intrinsic hyperfine spin), we can find the energy spectrum for any $\alpha$ by projecting the Hamiltonian ${\rm H}_0$ onto the ($n, \ell$) subspace. Here, $n$ and $\ell = j + m_F$ are the radial and orbital quantum numbers for the oscillator. This gives a tri-diagonal matrix, which can be numerically diagonalized by incorporating enough $3$D oscillator levels~\cite{anderson_three-dimensional_2013}. See~\cite{supplementary} for details. In Fig.~\ref{fig:GS_spectrum} we provide different energy curves as a function of $\alpha$, for different $n$ and $j$ values.

More importantly, from the point of view of the adiabatic flow of local spin, at large $\alpha$ the radial part of the scalar function $\psi$ approaches the $1$D harmonic oscillator centered at radius $r_0$ (c.f. Eq.~\eqref{heff}), while the angular structure is dictated by the Hamiltonian $H_{\Omega}= [({\bm r}\times({\bm p}+{\bm A}))^2 + F]/(2\alpha^2F^2)$. 
That is, $\psi({\bm r}) \rightarrow r^{-1}h_{\tilde{n}}(r-r_0)\,g^{m}_{F,j}(\theta,\varphi)$, where $h_{\tilde{n}}$ are the $1$D harmonic oscillator states and $g^{m}_{F,j}$ are the eigenstates of the angular Hamiltonian $H_{\Omega}$ (known as the monopole harmonics~\cite{WU1976365}). Then with the ansatz $f^{\ell}_{\tilde{n}jF}(r) \equiv r^{-1}h_{\tilde{n}}(r-r_0)\,\beta^{\ell}_{F,j}$ in Eq.~\eqref{eq:Psi_full_decompose}, the $\beta$ coefficients must be such that the following holds
\begin{align}
\label{eq:adiabatic_flow_condition}
    \sum^{j+F}_{\ell=|j-F|}\beta^{\ell}_{F,j}|\chi^{\ell F}_{jm}\rangle = g^{m}_{F,j}(\theta,\varphi)|F_{\hat{\bm r}}\rangle\,.
\end{align}
This is the radial spin flow correspondence (which holds for any $j \geq F$). Using this, the energy spectrum comes out to be~\cite{supplementary} 
\begin{align}
\label{eq:adiabatic_energy}
   E_{\tilde{n}jF} + \frac{\alpha^2F^2}{2} &\approx \tilde{n} + \frac{1}{2} + \frac{j(j+1)-F(F-1)}{2\alpha^2F^2}\,, 
\end{align}
where the last term is just the spherical Landau levels (LLs)~\cite{haldane_FractionalQuantizationHall_1983}, plus a shift $F/(2\alpha^2 F^2)$ owing to the `electric dipole moment'. We note that the radial spin flow correspondence~\eqref{eq:adiabatic_flow_condition}, without any reference to $H_{\Omega}$, fetches \textit{both} the $g$ functions and the $\beta$ coefficients, giving us all the LLs on the sphere~\cite{supplementary}. Our construction therefore reveals an alternative approach of constructing spherical LLs. It is clear from the energy spectrum that all the different $j$ levels approach one another as $\alpha$ increases, because the energies of different states get increasingly dominated by the Zeeman term. As an explicit example, consider the spin-$1$ case. For the $j=1$ level, we get $\beta^{\ell}_{1,1} = \{\sqrt{2},\sqrt{3},1\}/\sqrt{6}$  for $\ell = \{0,1,2\}$, and the following three degenerate states
\begin{align}
\label{lll}
    &\Bigl\{\Psi_{\tilde{n}1-11},\,\Psi_{\tilde{n}101},\,\Psi_{\tilde{n}111}\Bigr\} \approx \sqrt{\frac{3}{4\pi}}\,\frac{1}{r}h_{\tilde{n}}(r-\alpha)\,\times\nonumber\\
    & \quad\Biggl\{ \sin^{2}\left(\frac{\theta}{2}\right) e^{-i\varphi},\, \frac{\sin\theta}{\sqrt{2}},\, \cos^{2}\left(\frac{\theta}{2}\right)e^{i\varphi} \Biggr\} \left|{F}_{\hat{\bm r}} \right\rangle \,.
\end{align}
The inset in Fig.~\ref{fig:GS_spectrum} compares the energy with the Gross-Pitaevskii (GP) equation, and numerical diagonalization. It is evident that the radial spin flow correspondence holds well within $\sim 0.1\%$ for $\alpha \gtrsim 4$.

\subparagraph*{Creating vortices from different spin states:}
Conservation of ${\bf J}$ can be exploited to create vortex patterns/Dirac strings, as follows. Starting at $\alpha=0$, we can prepare the system in its ground state, carrying zero orbital angular momentum (OAM) $\langle{\bf L}\rangle_{\rm ini}=0$ and any desired spin configuration $|\zeta^x\rangle$ carrying spin angular momentum (SAM) $\langle{\bf F}\rangle_{\rm ini}$. Then $\alpha$ is increased adiabatically, and in the process some of the SAM gets transferred to the OAM while keeping the total $\langle{\bf J}\rangle$ fixed. At sufficiently large $\alpha$ owing to adiabatic spin flow, we converge to a vortex pattern in the final state carrying final OAM $\langle{\bf L}\rangle_{\rm fin}=\langle{\bf F}\rangle_{\rm ini} / (1+F)$, and SAM $\langle{\bf F}\rangle_{\rm fin}=F\langle{\bf F}\rangle_{\rm ini}/(1+F)$~\cite{supplementary}.

\begin{figure}[t]
    \centering
    \includegraphics[width=9cm]{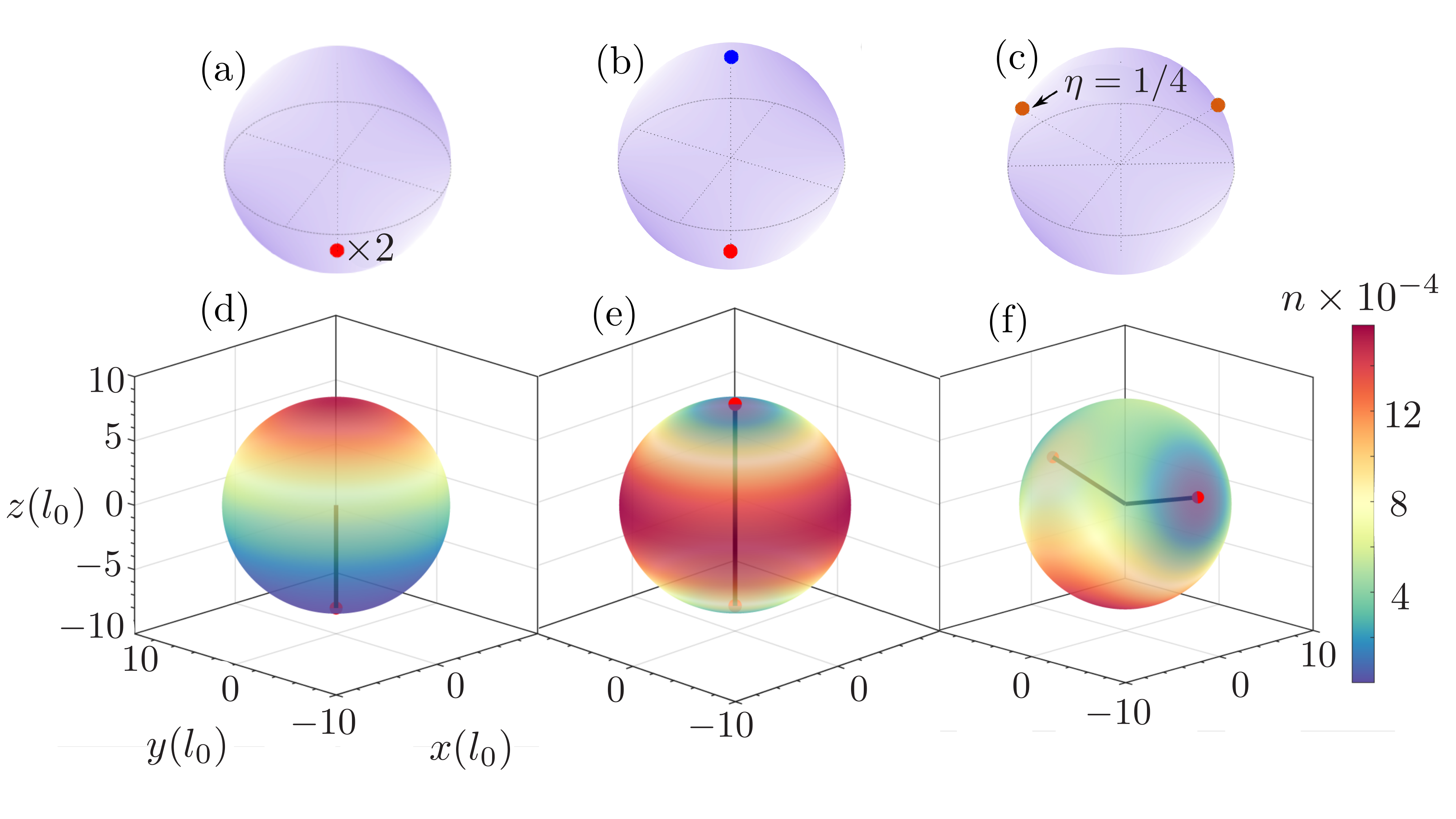}
    \caption{Upper panel: Majorana representations of spin-1 (a) ferromagnetic state $ \zeta^{F}=(1,0,0)^T $, (b) polar state $ \zeta^{P}=(0,1,0)^T $, and (c) mixed state $ \zeta^{\eta}=(\sqrt{\eta},0,\sqrt{1-\eta})^T $ with $\eta=1/4$ on the Bloch sphere.
    Lower panel: corresponding density profiles on the sphere of radius $r_0=\alpha=8$. Black lines represent the vortex lines, with the intersection points with atomic cloud at $r_0$ being Majorana stars.}
    \label{fig:majorana}
\end{figure}

We can also predict the orientation of these Dirac strings (which are lines of singularities) by considering the geometric/Bloch sphere representation for spin-$F$.  We note that at points where the strings/vortices intersect the sphere, the `wavefunction' must vanish. Then, owing to the transfer of initial SAM to final TAM during the adiabatic spin flow, these intersection points should be where the initial spin configuration $|\zeta^x\rangle$ was orthogonal to the final spin configuration $|F_{\hat{\bm r}}\rangle$. 
With $|\zeta^x\rangle = \sum_{m_F=-F}^F\zeta^x_{m_F}|m_{F\,{\hat{\bm z}}} \rangle$ and $|F_{\hat{\bm r}}\rangle = e^{-i \varphi F_{z}} e^{-i \theta F_{y}} \left|F_{\hat{\bm z}} \right\rangle$, this means those points $(\theta,\varphi)$ on the sphere where $\langle\zeta^x|F_{\hat{\bm r}}\rangle=0$:
\begin{align}
\label{MP}
    \sum^{2F}_{k=0} e^{-ik\varphi}\,\zeta^x_{F-k}  \sqrt{\begin{pmatrix}
        2F\\
        k
    \end{pmatrix}}
    \cos^{2F}\left(\frac{\theta}{2}\right)\,\tan^{k}\left(\frac{\theta}{2}\right)=0\,.
\end{align}
Note that these points are nothing but the so called Majorana stars, and our engineering of the Dirac strings reveals the connection between spin and real space. This connection is embodied in the $\mathrm{SO(3)}$ symmetry in both
the spinor Boson gas and the simulated monopole system. More explicitly,
the symmetries of $\left|\zeta^{x}\right\rangle $ correspond to the
operations under which the set of vortex locations $\left\{ (\theta_{i},\varphi_{i})\right\} $
on the Haldane sphere are invariants.

As explicit examples, consider the following ferromagnetic, polar and mixed states to begin with: $|\zeta^{F}\rangle=(1,0,0)^T$, $|\zeta^{P}\rangle=(0,1,0)^T$, and $|\zeta^{\eta}\rangle=(\sqrt{\eta},0,\sqrt{1-\eta})^T$. Based on our discussion above and Eq.~\eqref{MP}, each would correspond to two Dirac strings/vortices at large $\alpha$, originating from the origin. Their locations are calculated to be $\theta = \pi,\pi$ for ferromagnetic, $\theta = 0,\pi$ for polar, and $(\theta,\varphi)=(2\arctan[\eta/(1-\eta)]^{1/4},\pi \pm \pi/2)$ for the mixed state. The full final states, written in terms of the lowest LL (LLL) wavefunction in~\eqref{lll} are: $\Psi^F = \Psi_{0111}$, $\Psi^P = \Psi_{0101}$, and $\Psi^\eta = \sqrt{\eta}\,\Psi_{0111}+\sqrt{1-\eta}\,\Psi_{01-11}$. In the lower panel of Fig.~\ref{fig:majorana}, we show these states obtained using the normalized gradient flow method of~\cite{bao_ComputingGroundStates_2008}. The upper panel displays the Majorana representations of $\zeta^F$, $\zeta^P$ and $ \zeta^{\eta}$, respectively, where the highlighted points on the sphere correspond to~\eqref{MP}.

\begin{figure}[t]
    \centering
    \includegraphics[width=0.48\textwidth]{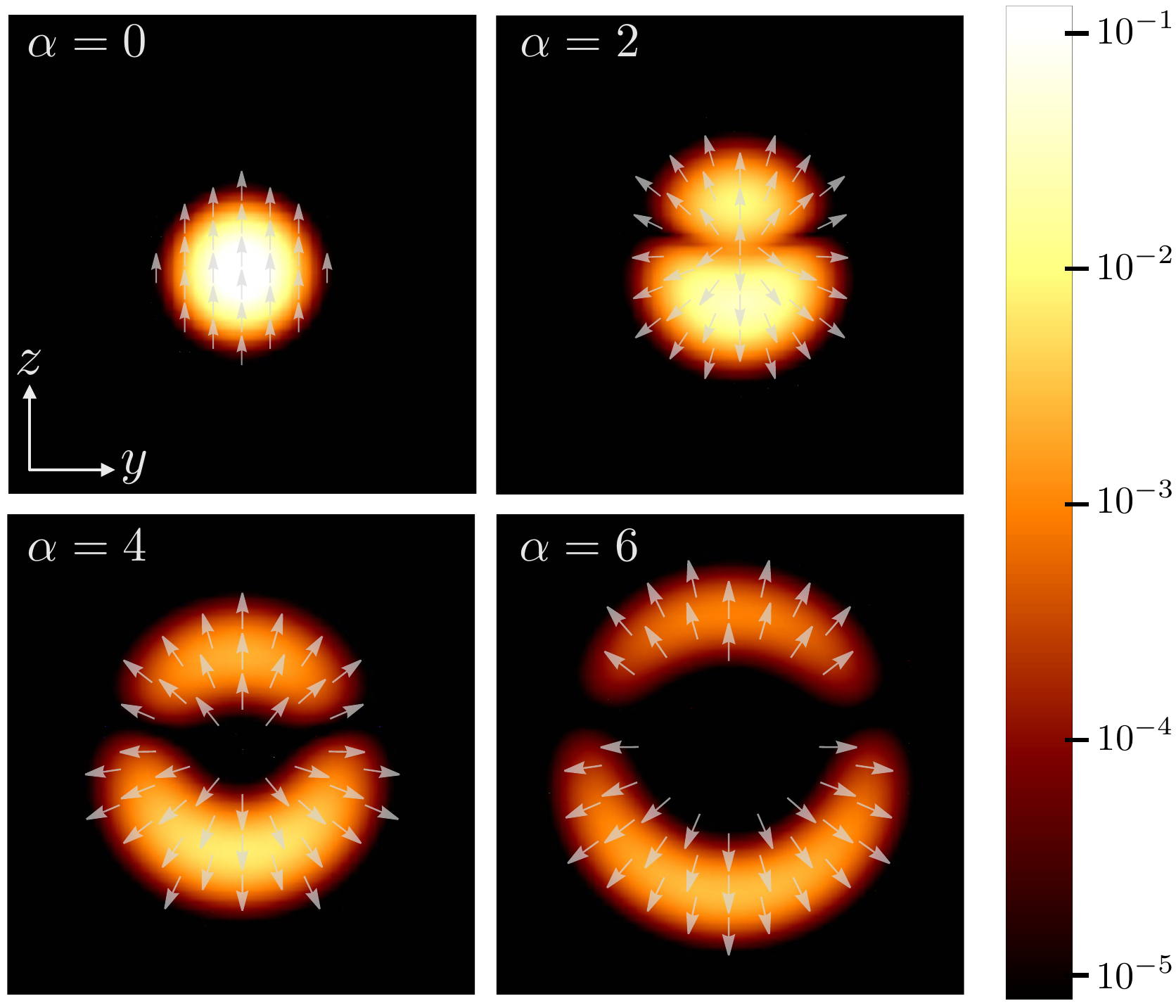}
    \caption{Evolution of the mixed state $\zeta^{\eta=1/4}$ when the field strength $\alpha$ is increased adiabatically from 0 to 6. The density profile (background color) and local spin expectation vector $\langle {\bf F}(\bm{r})\rangle$ (arrows) in the $x=0$ plane are plotted at four different times with the instataneous values of $\alpha$ indicated in the plots. For each plot, the length of the box along each direction is 25, and the calculation is done with a grid size $N^3=71^3$. A simulation animation is available \href{https://youtu.be/PCQGMqd-DfQ}{here}.}
    \label{fig:adiabatic}
\end{figure}

Using the integrator i-SPin 2~\cite{Jain:2023qty}, in Fig.~\ref{fig:adiabatic} we show the real time implementation of our idea, for the mixed state $|\zeta^{\eta=1/4}\rangle$. Starting with the 3D harmonic oscillator ground state dressed with the spin texture $|\zeta^{\eta=1/4}\rangle$, we adiabatically increase $\alpha$ from $0$ to $6$~\cite{supplementary}. As $\alpha$ increases, the initial mass density at the origin is pushed outwards, with the spin density aligning radially outwards. At large enough $\alpha$, two vortices intersecting the atomic cloud at $(\theta, \varphi) = (2\arctan[1/3]^{1/4},\pi\pm\pi/2)$, become apparent.  One can also get these locations by means of the vortex Dirac string connection: Rewriting the state as $\Psi^\eta = |\psi^\eta| |\tilde{F}_{\hat{\bm r}} \rangle$ where $|\tilde{F}_{\hat{\bm r}} \rangle = \exp( i\,{\rm arg}\,\psi^\eta )|F_{\hat{\bm r}} \rangle$, it can be shown that the effective gauge potential $\tilde{\bm A}=i\langle \tilde{F}_{\hat{\bm r}}|\nabla \tilde{F}_{\hat{\bm r}} \rangle$ contains two singularities located at $(2\tan^{-1}[1/3]^{1/4},\pi\pm\pi/2)$ when $\eta = 1/4$.

To summarize, we have established a one-to-one mapping between the spinor state and the vortex state on a sphere.\\

\begin{figure}[t]
    \centering
    \includegraphics[width=8.6cm]{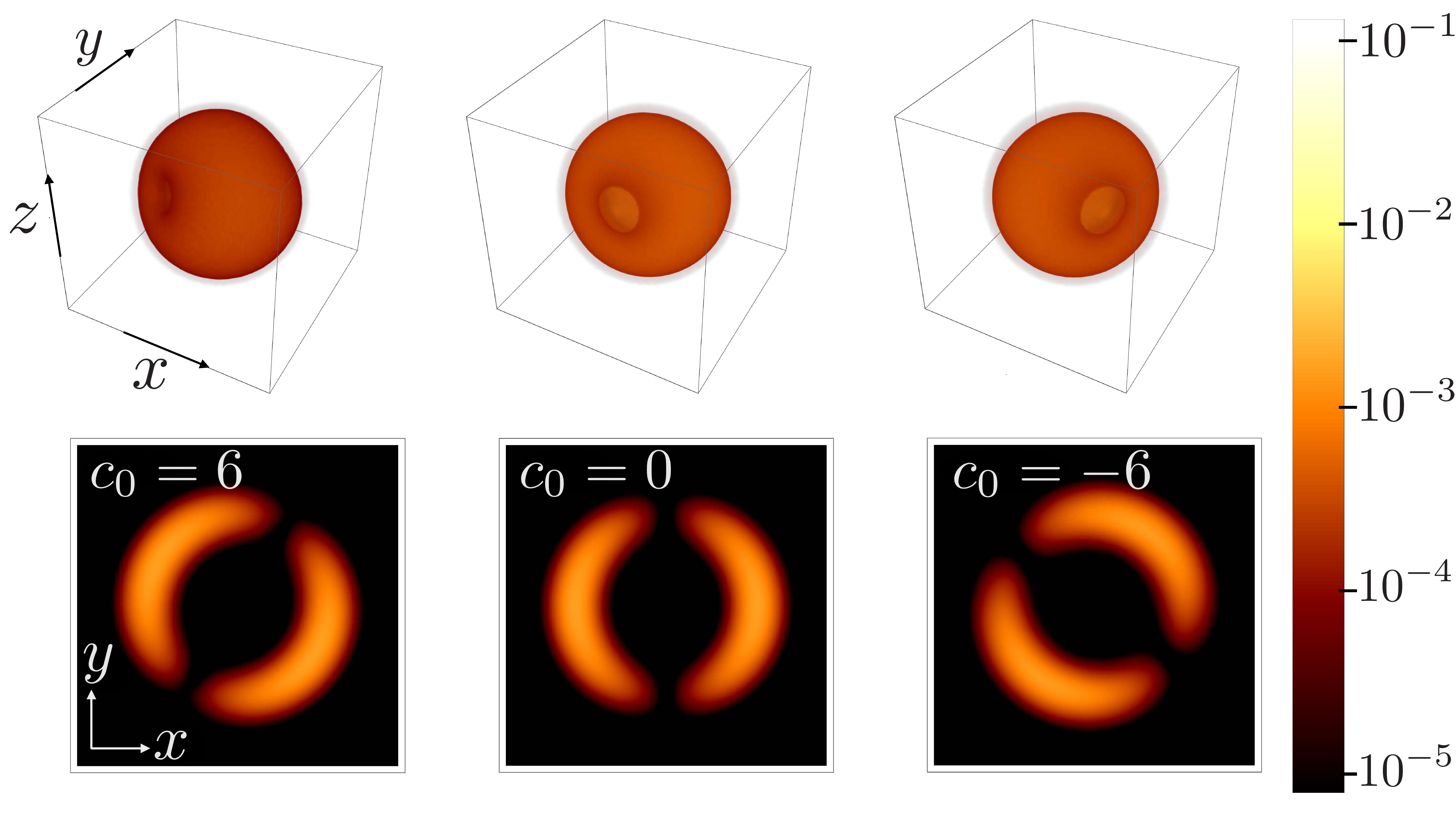}
    \caption{Snapshots of the mixed state atomic cloud for $\eta=1/4$ at $\alpha = 6$, for the three cases of attractive interaction ($c_0=-6$ and $c_2=-0.1$, left column), no interaction ($c_0=c_2=0$, middle column), and repulsive interaction ($c_0=6$ and $c_2=0.1$, right column), respectively. Other parameters are the same as in Fig.~\ref{fig:adiabatic}. The upper panel shows a 3D view, whereas the bottom panel shows the slice in the plane where the vortices intersect the atomic cloud (at polar angle $\theta=2\tan^{-1}(1/3)^{1/4} \approx 74.46$ degrees). Our results also confirmed that for the values of $c_2$ used here, the effect of the spin-dependent interaction is negligible.}
    \label{fig:effect_c0}
\end{figure}

\paragraph*{Effects of Interaction and Experimental Feasibility ---} Now we briefly discuss the effects of the mean-field interaction. The associated energy functional is $E_{\mathrm{int}}=(c_0/2)\int d^{3}{\bm r}\,  n^2({\bm r}) + (c_2/2)\int d^{3}{\bm r}~ {\bm F}({\bm r})\cdot {\bm F}({\bm r})$, where ${\bm F}({\bm r})$ is the local spin density, and the first and the second terms correspond to the spin-independent and -dependent interaction, respectively. Typically the two interaction strengths satisfy $|c_0| \gg |c_2|$. Figure~\ref{fig:effect_c0} illustrates the final density profiles for the mixed state (with $\eta=1/4$), for attractive, non-interacting and repulsive interaction cases, when starting from the respective ground states in the absence of the spin-dependent interaction we ramp up $\alpha$ from $0$ to $6$. While some of the qualitative physics remains the same, there are some important distinctions worth pointing out: (1) Without interactions, it can be seen that all single-particle states $\mathcal{Y}^m_{F,F}$ (or linear combinations thereof), are left invariant under SO($2$)$_{L_z+\phi}$ (here $\phi$ corresponds to gauge transformations). Including the interaction breaks this invariance and the states get rotated along the $z$ axis, as reflected by the rotation of vortex pairs. (2) In the final state, the angular momenta are not evenly distributed in the spin and the orbital sector. This is because the spin texture becomes more (an-)isotropic under the (attractive) repulsive interaction~\footnote{For the cases in Fig.~\ref{fig:effect_c0}, we have $\langle S_z \rangle \approx 0.22$ and $\langle L_z \rangle \approx 0.28$ for the repulsive case, while $\langle S_z \rangle \approx 0.28$ and $\langle L_z \rangle \approx 0.22$ for the attractive case. 
In comparison, $\langle S_z \rangle =\langle L_z \rangle = 0.25$ for the non-interacting/single-particle case.}.

For $N=10^4$ $^{87}\mathrm{Rb}$ atoms ($F=1$) in a trap with frequency $\omega=2\pi\times 100$ Hz, and Zeeman field $B_1\boldsymbol{r}$ with strength $B_1=1$ G cm$^{-1}$, $\alpha = r_0 \simeq 6$ is large enough where the spin flow correspondence holds well.
Furthermore, the energy scale of the contact interaction (per particle) is $\mathcal{E}_{c_0} \simeq c_0/(8 \pi \alpha^{2}) \approx 0.7$, and is only comparable with the corresponding LL gap $\Delta E_{\rm LL} \approx 1$. Therefore in such an experimental setup, the analytical results presented for non-interacting system remain qualitatively correct~\cite{supplementary}.\\

\paragraph*{Conclusions ---} The simulation of singular monopole potentials and the concomitant correspondence between the spin and the real space is a new feature of the spinor system under hedgehog magnetic field. The key ingredient is the rotational invariance of the Zeeman term ${\bm r}\cdot {\bf F}$. These will persist even when the $S_2$ manifold is deformed, as long as ${\rm J}_z$ remains conserved. Due to this rationale, we can investigate such features in other SO($3$) systems, such as the isotropic spin-orbital-coupling term ${\bm p}\cdot{\bf F}$, where bent vortex lines in solitons have been discovered~\cite{zhang_StableSolitonsThree_2015}. Remarkably, the correspondence presented in  this paper allows us to reveal  symmetries within the internal degrees of freedom, as manifesting in coordinate space.

In this work, we have found the spin real correspondence in the LLLs. Since stronger interactions may make the atoms occupy higher LLLs~\cite{greiter_LandauLevelQuantization_2011}, correspondence in these states may be found. The investigation in Haldane's spherical geometry was originally proposed for the study of the fractional quantum Hall effect.  Thus, fertile vortex configurations may enrich the exploration of many-body quantum matter in curved spatial geometry~\cite{lundblad_Shellpotentialsmicrogravity_2019,chakraborty_toroidaltrapcold_2016,turner_Vorticescurvedsurfaces_2010,ho_SpinorCondensatesCylindrical_2015,guenther_Quantizedsuperfluidvortex_2017,guenther_Superfluidvortexdynamics_2020,massignan_Superfluidvortexdynamics_2019,padavic_Vortexantivortexphysicsshellshaped_2020}.

\begin{acknowledgments}
This work was funded by National Natural Science Foundation of China (Grants No. 11974334,
No. 11774332) and Innovation Program for Quantum Science and Technology (Grant No. 2021ZD0301900). MA \& MJ were partly supported by DOE grant DE-SC0021619, and MJ is currently supported by a Leverhulme Trust Research Project (RPG-2022-145). H. P. is supported by the US NSF and the Welch Foundation (Grant No. C-1669).

\end{acknowledgments}

\bibliography{ref_spinor_BEC_monopole}

\newpage

\onecolumngrid

\appendix

\section{Constructing Landau levels and monopole harmonics using radial spin flow correspondence}

As discussed in the main text, adiabatic spin flow gives the following correspondence
\begin{align}
\label{eq:adiabatic_flow_condition_app}
    \sum^{j+F}_{\ell = |j-F|}\beta^{\ell}_{F,j}|\chi^{\ell F}_{jm}\rangle = g^{m}_{F,j}(\theta,\varphi)|F_{\hat{\bm r}}\rangle\,,
\end{align}
where $j \geq F$ (c.f. Eq.~\eqref{eq:adiabatic_flow_condition}). Since the sum over $\ell$ runs from $|j-F|$ to $j+F$, for any $j \geq F$ we have $2F+1$ $\beta^{\ell}_{F,j}$ coefficients to determine (which are independent of $m$). Also since $-j \leq m \leq j$,  we have $2j+1$ $g^m_{F,j}$ functions to determine. So we have $(2F+1)(2j+1)$ unknowns. Now since the above equation is a set of $2F+1$ equations (as $m_F \in [-F,F]$) for any $m \in [-j,j]$, we have $(2F+1)(2j+1)$ equations in total as well. So, we have a deterministic system and both the $\beta$ coefficients and the $g$ functions (which will turn out to be monopole harmonics) can be determined. This, for any $j \geq F$, will thus fetch the spherical Landau levels (LLs). While there are multiple ways to construct LLs, here we present an explicit approach to first see that $g^{m}_{F,j}$ are nothing but monopole harmonics $\mathcal{Y}^{m}_{q=F,j}$ (i.e. with monopole charge $q=F$ and total angular momentum $j$), which then can be used to obtain the $\beta$ coefficients.

From the left hand side of the above radial spin correspondence, which is an eigenstate of ${\bf J}^2$, we note that the right hand side must be too. Using ${\bf J} = {\bf L} + {\bf F}$ to operate on the right hand side, where $|F_{\hat{\bm r}}\rangle = e^{-i\varphi{\rm F}_{\hat{\bm z}}}\,e^{-i\theta{\rm F}_{\hat{\bm y}}}$, we get the following PDE after some algebraic manipulations 
\begin{align}
\label{eq:PDE_g}
    \Biggl[-\frac{1}{\sin\theta}\frac{\partial}{\partial\theta}\left(\sin\theta\frac{\partial}{\partial\theta}\right) - \frac{1}{\sin^2\theta}\frac{\partial^2}{\partial\varphi^2} + 2iF\frac{\cos\theta}{\sin^2\theta}\frac{\partial}{\partial\varphi} + F^2\frac{1}{\sin^2\theta}\Biggr]g^{m}_{F,j} = j(j+1)g^{m}_{F,j}\,.
\end{align}
This is exactly the PDE for monopole harmonics, with monopole charge $F$ and total angular momentum $j$~\cite{WU1976365}. That is
\begin{align}
    g^{m}_{F,j}(\hat{\bm r}) = \mathcal{Y}^{m}_{F,j}(\hat{\bm r})\,.
\end{align}
To get to~\eqref{eq:PDE_g} using ${\bf J}^2 = ({\bf L}+{\bf F})^2$, we used spherical forms for both ${\bf L}$ and ${\bf F}$. That is, ${\bf L} = i((\hat{\bm \theta}/\sin\theta)\,\partial_{\varphi} - \hat{\bm \varphi}\,\partial_{\theta})$ and ${\bf F} = \hat{\bm r}\,{\rm F}_{\hat{\bm r}} + \hat{\bm \theta}\,{\rm F}_{\hat{\bm \theta}} + \hat{\bm \varphi}\,{\rm F}_{\hat{\bm \varphi}}$, where ${\rm F}_{\hat{\bm r}} = \sin\theta\,\cos\varphi\,{\rm F}_{\hat{\bm x}} + \sin\theta\,\sin\varphi\,{\rm F}_{\hat{\bm y}} + \cos\theta\,{\rm F}_{\hat{\bm z}}$, ${\rm F}_{\hat{\bm \theta}} = \cos\theta\,\cos\varphi\,{\rm F}_{\hat{\bm x}} + \cos\theta\,\sin\varphi\,{\rm F}_{\hat{\bm y}} - \sin\theta\,{\rm F}_{\hat{\bm z}}$, and ${\rm F}_{\hat{\bm \varphi}} = -\sin\varphi\,{\rm F}_{\hat{\bm x}} + \cos\varphi\,{\rm F}_{\hat{\bm y}}$. We also used rotation relationships for spin matrices such as $e^{-i\varphi{\rm F}_{\hat{\bm z}}}\,{\rm F}_{\hat{\bm y}}\,e^{i\varphi{\rm F}_{\hat{\bm z}}} = 
-\sin\varphi\,{\rm F}_{\hat{\bm x}} + \cos\varphi\,{\rm F}_{\hat{\bm y}} = {\rm F_{\hat{\bm \varphi}}}$, and also eigen-equations ${\rm F}_{\hat{\bm r}}|F_{\hat{\bm r}}\rangle = F|F_{\hat{\bm r}}\rangle$, and ${\bf F}^2|F_{\hat{\bm r}}\rangle = F(F+1)|F_{\hat{\bm r}}\rangle$.

With the above correspondence with the monopole harmonics established, we can obtain the $\beta$ coefficients by simply taking the inner product of the right hand side of Eq.~\eqref{eq:adiabatic_flow_condition_app} with spinor harmonics $|\chi^{\ell F}_{j m}\rangle$:
\begin{align}
    \beta^{\ell}_{F,j} = \int\mathrm{d}\Omega\,g^{m}_{F,j}\langle \chi^{\ell F}_{jm}|F_{\hat{\bm r}}\rangle = \int\mathrm{d}\Omega\,\mathcal{Y}^{m}_{F,j}\langle \chi^{\ell F}_{jm}|F_{\hat{\bm r}}\rangle\,.
\end{align}

As explicit examples, for $F = 1$ we get $\beta^{\ell=\{0,1,2\}}_{1,1} = \{\sqrt{2},\sqrt{3},1\}/\sqrt{6}$ with
\begin{align}
    g^{-1}_{11} = \sqrt{\frac{3}{4\pi}}\,\sin^{2}(\theta/2)\,e^{-i\varphi}, \qquad g^{0}_{11} = \sqrt{\frac{3}{8\pi}}\,\sin\theta, \qquad 
    g^{1}_{11} = \sqrt{\frac{3}{4\pi}}\,\cos^{2}(\theta/2)\,e^{i\varphi}\,
\end{align}
for the lowest LLs ($j=F=1$). For the next LLs ($j = F+1 = 2$), we get $\beta^{\ell=\{1,2,3\}}_{1,2} = \{\sqrt{3},\sqrt{5},\sqrt{2}\}/\sqrt{10}$ with
\begin{align}
    g^{-2}_{12} &= \sqrt{\frac{5}{4\pi}}\,\sin^{2}(\theta/2)\sin\theta\,e^{-2i\varphi}, \qquad g^{0}_{12} = \sqrt{\frac{15}{32\pi}}\,\sin 2\theta, \qquad 
    g^{2}_{12} = -\sqrt{\frac{5}{4\pi}}\,\cos^{2}(\theta/2)\sin\theta\,e^{2i\varphi}\,\nonumber\\
    &\quad g^{-1}_{12} = \sqrt{\frac{5}{16\pi}}(\cos\theta - \cos 2\theta)\,e^{-i\varphi},\qquad g^{1}_{12} = \sqrt{\frac{5}{16\pi}}(\cos\theta + \cos 2\theta)\,e^{i\varphi}\,.
\end{align}

Similarly for $F = 2$, we get $\beta^{\ell=\{0,1,2,3,4\}}_{2,2} = \Bigl\{\sqrt{14}, \sqrt{28}, \sqrt{20}, \sqrt{7}, 1\Bigr\}/\sqrt{70}$ with
\begin{align}
    g^{-2}_{22} &= \sqrt{\frac{5}{4\pi}}\,\sin^{4}(\theta/2)\,e^{-2i\varphi}, \qquad g^{0}_{22} = \sqrt{\frac{15}{32\pi}}\,\sin^2\theta, \qquad 
    g^{2}_{22} = \sqrt{\frac{5}{4\pi}}\,\cos^{4}(\theta/2)\,e^{2i\varphi}\,\nonumber\\
    &\quad g^{-1}_{22} = \sqrt{\frac{5}{\pi}}\,\sin^{3}(\theta/2)\cos(\theta/2)\,e^{-i\varphi},\qquad g^{1}_{22} = \sqrt{\frac{5}{\pi}}\,\cos^{3}(\theta/2)\sin(\theta/2)\,e^{i\varphi}\,
\end{align}
for the lowest LLs ($j = F = 2$). For the next LLs ($j = F+1 = 3$), we get $\beta^{\ell=\{1,2,3,4,5\}}_{2,3} = \{\sqrt{6},\sqrt{15},\sqrt{14},\sqrt{6},1\}/\sqrt{42}$ with
\begin{align}
    g^{-3}_{23} &= \sqrt{\frac{21}{8\pi}}\,\sin^{4}(\theta/2)\sin\theta\,e^{-3i\varphi}, \qquad g^{0}_{23} = \sqrt{\frac{105}{32\pi}}\,\cos\theta\,\sin^2\theta, \qquad 
    g^{3}_{23} = -\sqrt{\frac{21}{8\pi}}\,\cos^{4}(\theta/2)\sin\theta\,e^{3i\varphi}\,\nonumber\\
    &\quad g^{-2}_{23} = \sqrt{\frac{7}{4\pi}}\,\sin^{4}(\theta/2)(3\cos\theta + 2)\,e^{-2i\varphi},\qquad g^{2}_{23} = \sqrt{\frac{7}{4\pi}}\,\cos^{4}(\theta/2)(3\cos\theta - 2)\,e^{2i\varphi}\,\nonumber\\
    &\quad g^{-1}_{23} = \sqrt{\frac{35}{8\pi}}\,\sin^{3}(\theta/2)\cos(\theta/2)\,(3\cos\theta + 1)\,e^{-i\varphi},\qquad g^{1}_{23} = \sqrt{\frac{35}{8\pi}}\,\cos^{3}(\theta/2)\sin(\theta/2)\,(3\cos\theta - 1)\,e^{i\varphi}\,.
\end{align}
And so on. It can also be noted that in general, $g^{-m}_{F,j}(\theta,\varphi) = (-1)^{j-m}\,g^{m}_{F,j}(\theta\rightarrow\theta+\pi,\varphi\rightarrow-\varphi)$.

\section{Calculations in the adiabatic regime}

In this section, we provide details of our calculations of the energy spectrum, and ratio of orbital to spin angular momentum, in the adiabatic regime.

\subsection{Energy spectrum}

In order to get the energy spectrum, we can take the expectation value of the Hamiltonian~\eqref{ham} with respect to the energy levels $\Psi_{njmF}$. Rewriting the Hamiltonian in spherical coordinates together with a shift by $\alpha^2F^2/2$ (in order to complete the square on the right hand side), we get
\begin{align}
\label{eq:Hamiltonian_main}
    H_0 + \frac{\alpha^2F^2}{2} = -\frac{1}{2r^2}\partial_r(r^2\partial_r) + \frac{1}{2}(r-\alpha F)^2 - \alpha r(\hat{\bm r}\cdot{\bf F} - F) + \frac{1}{2r^2}{\bf L}^2\,.
\end{align}
Then taking its expectation with respect to the state
\begin{align}
\label{eq:Psi_adiabatic_ansatz}
    \Psi_{njmF}({\bm r}) = \frac{1}{r}h_n(r-\alpha F)g^m_{F,j}(\theta,\varphi)|F_{\hat{\bm r}}\rangle\,,
\end{align}
where $h_n(r-\alpha F)$ are the 1D oscillator (centered at $r = \alpha F$) levels, $g^m_{F,j}$ are the monopole harmonics $\mathcal{Y}^m_{F,j}$, and $|F_{\hat{\bm r}}\rangle$ is the radially outward pointing spin state, gives
\begin{align}
\label{eq:Energy_adiabatic_1}
    E_{njF} + \frac{\alpha^2F^2}{2} = n + \frac{1}{2} + \frac{\langle{\bf L}^2\rangle}{2\alpha^2F^2}\,.
\end{align}
To get to the above, we have also used $\langle\hat{\bm r}\cdot{\bf F}\rangle = F$.
Now since the energy eigenstates are also eigenstates of both ${\bf J}^2$ and ${\bf F}^2$ with eigenvalues $j(j+1)$ and $F(F+1)$ respectively, we have
\begin{align}
\label{eq:Jsquared_exp}
    \langle{\bf J}^2\rangle = j(j+1) = \langle{\bf L}^2\rangle + F(F+1) + 2\langle{\bf F}\cdot{\bf L}\rangle\,.
\end{align}
To evaluate $\langle{\bf F}\cdot{\bf L}\rangle$, note that it is given by
\begin{align}
    \langle{\bf F}\cdot{\bf L}\rangle = \int\mathrm{d}\Omega\,g^{m\ast}_{F,j}\Bigl(\langle F_{\hat{\bm r}}|{\bf F}|F_{\hat{\bm r}}\rangle\,\cdot({\bf L}g^m_{F,j}) + g^m_{F,j}\langle F_{\hat{\bm r}}|{\bf F}\cdot({\bf L}|F_{\hat{\bm r}}\rangle)\Bigr)\,.
\end{align}
In the above, the first term is trivially zero since average spin points in the radial direction, while angular momentum lies along the sphere. For the second term, let us write $|F_{\hat{\bm r}}\rangle = \mathcal{M}|F_{\hat{\bm z}}\rangle$ (where $\mathcal{M} = e^{-i\varphi {\rm F}_{\hat{\bm z}}}e^{-i\theta {\rm F}_{\hat{\bm y}}}$ is the rotation operator), and now we are interested in the operator $\mathcal{M}^{\dagger}({\bf F}\cdot{\bf L})\mathcal{M}$. Using ${\bf L} = i\left(\frac{\hat{\bm \theta}}{\sin\theta}\frac{\partial}{\partial\varphi} - \hat{\bm \varphi}\frac{\partial}{\partial\theta}\right)$, writing everything in terms of Cartesian variables, and using rotation identities like $e^{-i\alpha {\rm F}_{\hat{\bm y}}}\,{\rm F}_{\hat{\bm z}}\,e^{i\alpha {\rm F}_{\hat{\bm y}}} = {\rm F}_{\hat{\bm z}}\cos\alpha + {\rm F}_{\hat{\bm x}}\sin\alpha$ (and similarly for rotations along other axes), we get $\mathcal{M}^{\dagger}({\bf F}\cdot{\bf L})\mathcal{M} = {\rm F}^2_{\hat{\bm z}} - {\rm F}^2 + {\rm F}_{\hat{\bm x}}{\rm F}_{\hat{\bm z}}\,\cot\theta$ after some algebra. Only the first two terms contribute towards the expectation with respect to the state $|F_{\hat{\bm z}}\rangle$, giving $F^2-F(F+1) = -F$. Since $g^m_{F,j}$ were normalized by definition, we get $\langle{\bf F}\cdot{\bf L}\rangle = -F$. Using this together with~\eqref{eq:Jsquared_exp} in~\eqref{eq:Energy_adiabatic_1}, fetches the desired energy levels, Eq.~\eqref{eq:adiabatic_energy}.

\subsection{Ratio of orbital to spin angular momentum}

In our adiabatic evolution, we are confined within the lowest energy states, i.e. the $j=F$ lowest Landau levels. With arbitrary initial spin dressing, the final state takes the general form
\begin{align}
\label{eq:Psi_adiabatic_ansatz_general}
    \Psi_{nFF}({\bm r}) = \frac{1}{r}h_{n}(r-\alpha F)\sum^{F}_{m=-F}c_m\,\mathcal{Y}^m_{q=F,j=F}(\theta,\varphi)|F_{\hat{\bm r}}\rangle\,\qquad{\rm with}\qquad \sum^{F}_{m=-F}|c_m|^2 = 1\,,
\end{align}
which is the same as~\eqref{eq:Psi_adiabatic_ansatz} with $g^m_{F,j=F}(\theta,\varphi)$ replaced by a sum over monopole harmonics with different magnetic quantum number $m$. With this and $\langle F_{\hat{\bm r}}|{\bf F}|F_{\hat{\bm r}}\rangle = F\hat{\bm r}$, we have for the full expectation
\begin{align}
\label{eq:J_exp}
    \langle{\bf J}\rangle = \langle{\bf L}\rangle + F\langle\hat{\bm r}\rangle\,,
\end{align}
where 
\begin{align}
    \langle\hat{\bm r}\rangle = \sum_{m,m'}c^{\ast}_mc_{m'}\int\mathrm{d}\Omega\,\mathcal{Y}^{m\ast}_{F,F}\,\hat{\bm r}\,\mathcal{Y}^{m'}_{F,F}.
\end{align}
With the general form
\begin{align}
    \mathcal{Y}^{m}_{F,F}(\theta,\varphi) = \sqrt{\frac{2F+1}{4\pi}\begin{pmatrix}
    2F\\
    F+m
    \end{pmatrix}}\,\cos^{F+m}(\theta/2)\,\sin^{F-m}(\theta/2)\,e^{im\varphi}\,,
\end{align}
we have for $\hat{r}_{\pm} \equiv \hat{\bm r}\cdot\hat{\bm x} + i\,\hat{\bm r}\cdot\hat{\bm y} = \sin\theta\,e^{\pm i\varphi}$ and $\hat{r}_z = \hat{\bm r}\cdot\hat{\bm z} = \cos\theta$, the following
\begin{align}
    \langle\hat{r}_{\pm}\rangle &= \frac{1}{F+1}\sum^{F}_{m=-F}c^{\ast}_{m\pm 1}c_{m}\sqrt{F(F+1)-m(m\pm 1)} = \frac{1}{F+1}\langle{\rm J}_{\pm}\rangle\,,\nonumber\\
    \langle\cos\theta\rangle &= \frac{1}{F+1}\sum^{F}_{m=-F}m|c_m|^2 = \frac{1}{F+1}\langle{\rm J}_{z}\rangle\,.
\end{align}
The second equalities in each of the two expressions above is simply due to $\mathcal{Y}^m_{F,F}({\bm r})$ being eigenstates of $\{{\bf J}^2,{\rm J}_z\}$, and can be obtained straightforwardly using angular momentum operator algebra. Therefore, from~\eqref{eq:J_exp}, we have that \begin{align}
\label{eq:L_F_ratio}
    \langle{\bf L}\rangle = \frac{1}{F}\langle{\bf F}\rangle = \frac{1}{F+1}\langle{\bf J}\rangle\,.
\end{align}
In our adiabatic evolution, the total angular momentum ${\bf J}$ is conserved throughout, with all of it initially being in the spin sector, $\langle{\bf J}\rangle = \langle{\bf F}\rangle_{\rm ini}$. Eq.~\eqref{eq:L_F_ratio} then dictates that $F/(F+1)$ of the total initial spin remains in the spin sector, with the remaining $1/(F+1)$ is transferred to the orbital sector.

\section{Energy spectrum using the $\textit{su}$($1,1$)$\times$SO($3$) group algebra}
Following the steps of~\cite{anderson_three-dimensional_2013}, we can numerically diagonalize the Hamiltonian. To begin with, using creation and annihilation operators $a^{\dagger}_i$ and $a_i$ where $i = \{x,y,z\}$, we note that it is possible to get generators of both \textit{su}($1,1$)$_{\rm S}$ and SO($3$)$_{\rm L}$. Rewriting $a_\pm = \mp (a_{\hat{\bm x}} \mp ia_{\hat{\bm y}})/\sqrt{2}$ and $a_{\hat{\bm z}} = a_0$, we have the following
\begin{align}
{\rm S}_{+} & = \frac{1}{2}(a_{0}^{\dagger})^{2}-a_{+}^{\dagger}a_{-}^{\dagger},\qquad\qquad\qquad\qquad\qquad\qquad\qquad {\rm L}_{+} = \sqrt{2}\left(a_{+}^{\dagger}a_{0}+a_{-}a_{0}^{\dagger}\right)\,\\
{\rm S}_{-} & = \frac{1}{2}(a_{0})^{2}-a_{+}a_{-},\qquad\qquad\qquad\qquad\qquad\qquad\qquad {\rm L}_{-} = \sqrt{2}\left(a_{+}a_{0}^{\dagger} + a_{-}^{\dagger}a_{0}\right)\,\\
{\rm S}_{0} & =\frac{1}{2}\left(a^{\dagger}_{+}a_{+} + a^{\dagger}_{-}a_{-} + a^{\dagger}_{z}a_{0} +\frac{3}{2}\right),\,\quad\qquad\qquad\qquad {\rm L}_{0}=a_{+}^{\dagger}a_{+} - a_{-}^{\dagger}a_{-}\,.
\end{align}
It can be checked that
the set of S and L obey the desired Lie algebra of the \textit{su}(1,1)$_{\rm S}$ and SO($3$)$_{\rm L}$ respectively:
\begin{align}
\label{eq:SU11_SO3_commutator}
\left[{\rm S}_{+},{\rm S}_{-}\right]=-2{\rm S}_{0},\quad & {\rm and}\quad\left[{\rm S}_{0},{\rm S}_{\pm}\right]=\pm {\rm S}_{\pm}\nonumber\\
\left[{\rm L}_{+},{\rm L}_{-}\right]=2{\rm L}_{0},\quad & {\rm and}\quad\left[{\rm L}_{0},{\rm L}_{\pm}\right]=\pm {\rm L}_{\pm}\,,
\end{align}
with $[{\rm S}_{\mu},{\rm L}_{\nu}] = 0$. Using the fact that ${\bm r}\cdot{\bf L} = 0$ and ${\bf J} = {\bf F} + {\bf L}$, the full Hamiltonian can then be written as ${\rm H} = {\rm S}_0 - \alpha\,{\bm r}\cdot{\bf F} = {\rm S}_0 - \alpha\,{\bm r}\cdot{\bf J}$, where we further decompose the Zeeman term as
\begin{align}
    {\bm r}\cdot{\bf J} = \frac{1}{\sqrt{2}}\Biggl[\frac{{\rm J}_{+}\,a^{\dagger}_{-} - {\rm J}_{-}\,a^{\dagger}_{+}}{\sqrt{2}} + {\rm J_0}\,a^{\dagger}_{0} + {\rm h.c.}\Biggr] \equiv \frac{\mathcal{A}^{\dagger} + {\rm h.c.}}{\sqrt{2}}\,.
\end{align}
Here, ${\bm r} = ({\bm a} + {\bm a}^{\dagger})/\sqrt{2}$, ${\rm J}_{\pm} = {\rm J}_{\hat{\bm x}} \pm i{\rm J}_{\hat{\bm y}}$ and ${\rm J}_0 = {\rm J}_{\hat{\bm z}}$. With $\alpha = 0$, we simple have a $3$D oscillator with an intrinsic (hyperfine-)spin, the Hamiltonian for which is just ${\rm H} = {\rm S}_0$. The energy eigenstates are common eigenstates of the operators ${\rm S}_0$, the Casimir ${\rm S}^2 \equiv {\rm S}_0^2 - ({\rm S}_{+}{\rm S}_{-} + {\rm S}_{-}{\rm S}_{+})/2$, ${\rm L}^2$, and ${\rm L}_z$. These can be labelled by the quantum numbers $(n,\ell,m_{\ell})$, with eigenvalues $2n+\ell+3/2$. We shall denote them by $|n,\ell,m_{\ell}\rangle$, and can be obtained by repeated actions of ${\rm S}^{n}_{+}$ and ${\rm L}^{\ell-m_{\ell}}_{-}$, and $(a^{\dagger}_{+})^{\ell}$~\cite{anderson_three-dimensional_2013}. 

Including the $\alpha$ dependent Zeeman term, couples the different $3$D oscillator states. The quantum numbers associated with the full Hamiltonian are $(\tilde{n},j,m,F)$, and the eigenstates are some 
$n$ and $\ell$ superpositions (with $|j-F| \leq \ell \leq j+F$ as required by the triangle inequality) of the basis set
\begin{align}
    |n,j,m,F,\ell\rangle \equiv \sum^{F}_{m_F = -F}c^{\ell m_F}_{j m}|n,\ell,m-m_{F}\rangle\otimes|m_{F_{\hat{\bm z}}}\rangle\,.
\end{align}
Here $|n,\ell,m-m_{F}\rangle$ are the $3$D oscillator states as stated above. With this, and noting that ${\bf J}$ is conserved in the system, we wish to find the matrix elements of ${\bm r}\cdot{\bf J}$ for a fixed $(j,m)$. These should be independent of $m$ since ${\bf J}$ is conserved. To find these matrix elements, we first recapitulate the action of $a^{\dagger}_{\mu}$ on a state $|n,\ell,m_{\ell}\rangle$~\cite{anderson_three-dimensional_2013}: 
\begin{align}
\label{eq:adagger_matrix_element}
a_{\mu}^{\dagger}\left|n,\ell, m_{\ell}\right\rangle = b^{+}(n,\ell)\,d_{\mu}^{+}(\ell, m_{\ell})\left|n,\ell+1,m_{\ell}+\mu\right\rangle + b^{-}(n,\ell)\,d_{\mu}^{-}(\ell, m_{\ell})\left|n+1,\ell-1,m_{\ell}+\mu\right\rangle\,,
\end{align}
where 
\begin{align}
\label{eq:b_p_m}
b^{+}(n,\ell)&=\sqrt{\frac{n+\ell+3/2}{(\ell+3/2)(\ell+1/2)}}\,\nonumber\\
b^{-}(n,\ell)&=\sqrt{\frac{n+1}{(\ell+1/2)(\ell-1/2)}}\,,\nonumber\\
d_{\mu}^{+}(\ell, m_{\ell})&=\left(\frac{1}{\sqrt{2}}\right)^{1+\left|\mu\right|}
\begin{cases}
\sqrt{(l+m_{\ell}+2)(l+m_{\ell}+1)} & \mu=+1\nonumber\\
\sqrt{(l+m_{\ell}+1)(l-m_{\ell}+1)} & \mu=0\nonumber\\
\sqrt{(l-m_{\ell}+2)(l-m_{\ell}+1)} & \mu=-1.
\end{cases},\nonumber\\
d_{\mu}^{-}(\ell,m_{\ell})&=(-1)^{\mu}\left(\frac{1}{\sqrt{2}}\right)^{1+\left|\mu\right|}
\begin{cases}
\sqrt{(l+m_{\ell})(l+m_{\ell}-1)} & \mu=-1\nonumber\\
\sqrt{(l+m_{\ell})(l-m_{\ell})} & \mu=0\nonumber\\
\sqrt{(l-m_{\ell})(l-m_{\ell}-1)} & \mu=+1.
\end{cases}\,. 
\end{align}
For our purposes, $m_{\ell} = m-m_F$ in the above. The above can be obtained using the various commutation relations between $a_{\mu}$ and ${\rm S}_{\mu}$, and $a_{\mu}$ and ${\rm L}_{\mu}$ (which can be obtained straightforwardly using the commutation relations~\eqref{eq:SU11_SO3_commutator}). The action of $a_{\mu}$ on $|n,\ell,m_{\ell}\rangle$ can be obtained in a similar fashion. Also for convenience/better illustration, $\mu = \pm \rightarrow \pm 1$ in the expressions for $d^{\pm}_{\mu}$. Next, the action of ${\bf J}$ on a state $|n,j,m,F,\ell\rangle$ is ${\rm J}_{\pm}|n,j,m,F,\ell\rangle = \sqrt{j(j+1)-m(m\pm 1)}\,|n,j,m\pm 1,F,\ell\rangle$ and ${\rm J}_{0}|n,j,m,F,\ell\rangle = m|n,j,m,F,\ell\rangle$. 

With the above, we can work out the matrix elements of $\mathcal{A}$ and $\mathcal{A}^{\dagger}$, in the subspace of fixed ($j,m$): 
\begin{align}
    [\mathcal{A}^{\dagger}(j,F)]_{n',n;\ell,\ell'} &\equiv \langle n',j,m,F,\ell'|\mathcal{A}^{\dagger}|n,j,m,F,\ell\rangle\nonumber\\
    & = b^{+}(n,\ell)\sum^{F}_{m_F=-F}c^{\ell+1, m_{F}\,\ast}_{jm}\,\mathcal{C}_{+}(j,m,\ell,m_F)\,\delta_{n',n}\,\delta_{\ell',\ell+1}\nonumber\\
    &\, + b^{-}(n,\ell)\sum^{F}_{m_F=-F}c^{\ell-1, m_{F}\,\ast}_{jm}\,\mathcal{C}_{-}(j,m,\ell,m_F)\,\delta_{n',n+1}\,\delta_{\ell',\ell-1}\,,
\end{align}
where
\begin{align}
\label{eq:C_coeff_new}
\mathcal{C}_{+}(j,m,\ell,m_F) &\equiv \frac{1}{2\sqrt{2}}\Bigl[\sqrt{j(j+1)-m(m+1)}\sqrt{(\ell-m+m_F+1)(\ell-m+m_F)}\,c^{\ell m_F}_{j,m+1}\nonumber\\
&\qquad\qquad - \sqrt{j(j + 1) - m(m - 1)}\sqrt{(\ell+m-m_F+1)(\ell+m-m_F)}\,c^{\ell m_F}_{j,m-1}\nonumber\\
&\qquad\qquad + 2m\sqrt{(\ell+m-m_F+1)(\ell-m+m_F+1)}\,c^{\ell m_F}_{j,m}\Bigr]\,\nonumber\\
\mathcal{C}_{-}(j,m,\ell,m_F) &\equiv \frac{1}{2\sqrt{2}}\Bigl[-\sqrt{j(j + 1) - m(m + 1)}\sqrt{(\ell+m-m_F+1)(\ell+m-m_F)}\,c^{\ell m_F}_{j,m+1}\nonumber\\
&\qquad\qquad + \sqrt{j(j + 1) - m(m - 1)}\sqrt{(\ell-m+m_F+1)(\ell-m+m_F)}\,c^{\ell m_F}_{j,m-1}\nonumber\\
&\qquad\qquad + 2m\sqrt{(\ell+m-m_F)(\ell-m+m_F)}\,c^{\ell m_F}_{j,m}\Bigr]\,.
\end{align}
Summing over $m_F$ results in the following explicit form
\begin{align}
    [\mathcal{A}^{\dagger}(j,F)]_{n',n;\ell',\ell} &= \Biggl[\frac{\sqrt{(j-\ell+F)(F-j+\ell+1)(j-F+\ell+1)(j+F+\ell+2)}}{2\sqrt{2}}\sqrt{\frac{n+\ell+3/2}{(\ell+3/2)(\ell+1/2)}}\Biggr]\delta_{n',n}\delta_{\ell',\ell+1}\nonumber\\
    &\, + \Biggl[\frac{\sqrt{(j+\ell-F)(F+j-\ell+1)(F-j+\ell)(F+j+\ell+1)}}{2\sqrt{2}}\sqrt{\frac{n+1}{(\ell+1/2)(\ell-1/2)}}\Biggr]\delta_{n',n+1}\delta_{\ell',\ell-1}\,,
\end{align}
and is indeed independent of $m$. The matrix elements of $\mathcal{A}$ are simply obtained through conjugation. Along with the triangle inequality $\left|j-F\right|\leq \ell \leq j+F$, the obtained matrix elements of $\mathcal{A}^{\dagger}$ and $\mathcal{A}$ fetch the following matrix elements for the Hamiltonian in $(n,\ell)$ subspace:
\begin{align}
    [H(j,F)]_{n',n;\ell',\ell} = \Biggl(2n+\ell + \frac{3}{2}\Biggr)\delta_{n',n}\,\delta_{\ell',\ell} -\frac{\alpha}{\sqrt{2}}\Bigl([\mathcal{A}^{\dagger}(j,F)]_{n',n;\ell',\ell} + {\rm h.c.}\Bigr)\,.
\end{align}
This is a tridiagonal matrix, and can be numerically diagonalized by including a lot of $3$D oscillator states. Using the triangle inequality to write $\ell = j + m_F$, we do this rather in the $(n,m_F)$ subspace, for $N = 2n + j + m_F \leq 100$. The spectrum for $F=1$, as a function of $\alpha$, is shown in Fig.~\ref{fig:GS_spectrum} of the main text.

\section{Numerical Analysis}

\subsection{Obtaining ground states by minimizing the energy functional}

With all the physical parameters intact, the Gross-Pitaevskii equation characterizing our $3$D
system is
\begin{align}
    i\hbar\frac{\partial}{\partial t}\Psi=\left[-\frac{\hbar^{2}}{2M}\nabla^{2}+\frac{1}{2}M\omega^{2}r^{2}-\mu_{B}g_{F}\,{\bm B}\cdot{\bf F}+\frac{1}{2}\bar{c}_{0}n\left({\bm r}\right)+\frac{1}{2}\bar{c}_{2}\left\langle {\bf F}\right\rangle \cdot{\bf F}\right]\Psi\,.
\end{align}
Here, $M$ is the mass of the particle/atom; $\omega$ is the external trap frequency; $\mu_B$ is the Bohr magneton; $g_F$ is the hyperfine g-factor of the atom; ${\bm B} = 2B_1\,{\bm r}$ is the effective magnetic field in the rotating frame, with $B_1$ being the amplitude parameter of the quadruple magnetic field in the lab frame~\cite{zhou_SyntheticLandauLevels_2018}; $\bar{c}_0 = 4\pi\hbar^2(a_2 + 2a_0)/3M$ and $\bar{c}_2 = 4\pi\hbar^2(a_2 - a_0)/M$ are effective $2$ body interaction parameters where $a_0$ and $a_2$ are the s-wave scattering lengths for the total spin equal to $0$ and $2$ channels; and the ``wave function" is normalized as $\int\mathrm{d}^3x\,\Psi^{\dagger}\Psi = N$ where $N$ is the total number of particles. With the re-scalings $t\to t/\omega$,
${\bm r}\to {\bm r}l_{s}$, and $\Psi\to\Psi\sqrt{N}l_{s}^{-3/2}$, where $l_{s}=\sqrt{\hbar/\left(M\omega\right)}$, we get the dimensionless GPE equation
\begin{equation}
\label{eq:dimensionless_GPE}
i\frac{\partial}{\partial t}\Psi =\left[-\frac{\nabla^{2}}{2}+\frac{r^{2}}{2}-\alpha\,{\bm r}\cdot{\bf F}+\frac{1}{2}c_{0}n\left({\bm r}\right)+\frac{1}{2}c_{2}\left\langle {\bf F}\right\rangle \cdot{\bf F}\right]\Psi\,,
\end{equation}
where $\alpha = 2\mu_{B}g_{F}B_1l_{s}/\left(\hbar\omega\right)=\omega_{B}/\omega$ with $\omega_{B}$ being the Larmor frequency $\omega_{B}=2g_{F}\mu_{B}B_1l_{s}/\hbar$,
and the interaction parameters are $c_0 = 4\pi N(a_2+2a_0)/3l_s$ and $c_1 = 4\pi N(a_2-a_0)/3l_s$. Two main conserved quantities, associated with Eq.~\eqref{eq:dimensionless_GPE} include
the total particle number $N$, 
and the total angular momentum
\begin{align}
\label{eq:totalAngularMomentum}
\left\langle {\bf J}\right\rangle = \left\langle {\bf L}\right\rangle +\left\langle{\bf F}\right\rangle = \int d^{3}{\bm r}\left[\sum_{m_{F}}\Psi_{m_{F}}^{*}{\bf L}\Psi_{m_{F}}+\sum_{m_{F}}\sum_{m_{F}^{\prime}}\Psi_{m_{F}}^{*}\left({\bf F}\right)_{m_{F}m_{F}^{\prime}}\Psi_{m_{F}^{\prime}}\right]\,.
\end{align}
The ground state $\Psi_{g}$ of the system can then be obtained by minimizing the energy
functional 
\begin{align}
\mathcal{E}\left[\Psi\left(\cdot,t\right)\right] =\int d^{3}{\bm r}\left\{ \sum_{m_{F}}\left(\frac{1}{2}\left|\nabla\Psi_{m_{F}}\right|^{2}+V\left(r\right)n_{m_{F}}\right)-\alpha{\bm r}\cdot\left\langle \boldsymbol{F}\right\rangle +\frac{1}{2}c_{0}n^{2}\left({\bm r}\right)+\frac{1}{2}c_{2}\left\langle \boldsymbol{F}\right\rangle ^{2}\right\}, \label{eq:energy_functional}
\end{align}
subjected to the two constraints of number and angular momentum conservation. This can be simply done by introducing Lagrange multipliers $\mu$ and ${\bm \lambda}$ for the $4$ conserved numbers, meaning one minimizes the following
\begin{align}
\mathcal{E}_{L}\left[\Psi\left(\cdot,t\right)\right] = \mathcal{E}\left[\Psi\left(\cdot,t\right)\right]-\mu N-{\bm \lambda}\cdot\left\langle {\bf J}\right\rangle\,,
\end{align}
using the continuous normalized gradient flow method (imaginary time evolution), as described in Ref.~\cite{bao_ComputingGroundStates_2008}. 
Setting both $c_0 = c_1 = 0$, in Fig.~\ref{fig:majorana} of the main text, we show the single particle ground states (for $F=1$) for the three different spin configurations considered there. 

\subsection{Full, real time simulations}

We have performed real time adiabatic flow simulations of our rescaled system~\eqref{eq:dimensionless_GPE}, to confirm and validate our results in this paper. We used the integrator i-SPin $2$~\cite{Jain:2023qty} developed by some of us, in order to perform these real time simulations. In general, the pseudo-spectral algorithm in i-SPin $2$ is time-reversible, along with norm and spin preserving to machine precision. It can also handle self-interactions as well as couplings to time-dependent external fields. For the interested reader, the details of the algorithm and numerical implementation can be found in that paper.

For our present purpose, we begin by constructing the ground state of the initial system (with $\alpha = 0$ and $c_2 = 0$). For $c_0 = 0$ this is simply a $3$D oscillator ground state (with an overall desired spin structure), whereas for $c_0 \neq 0$ it is not.  In general, we get to this state by using imaginary time evolution of the system (keeping the total particle number fixed, say $N = 1$). Taking this initial state, we then perform real time evolution of the system wherein $\alpha$ is increased from $0$ to some large number adiabatically. For this, we used a hyperbolic tangent function:
\begin{align}
    \alpha(t) = p_1\,\tanh[(t-t_0)/\tau] + p_2\,,
\end{align}
where the parameters $p_i$, $t_0$, and $\tau$ are chosen such that $\alpha(0) = 0$ and the final value approaches some desired number. For the simulations presented in the paper, we set $\alpha(t_f) = 6$. The time step used was $\Delta t \approx 0.04$ with $t_f \approx 107$. This meant $t_0 \approx 53.5$ and $\tau \approx 23$ in the above parameterization. Finally, the total box was a $71^3$ grid, with the length of the box in each direction being $25$. 

Starting with different ground states (with $c_0$ and different spin textures) with $c_2 = 0$, we have performed real time simulations for both the cases when $c_2$ was kept to zero, and was turned on to some small but finite value. In the main text, we show simulation results for the mixed state. Fig.~\ref{fig:adiabatic} shows the time evolution for the ``single particle" case ($c_0 = 0$). Fig.~\ref{fig:effect_c0} shows the same for $c_0 \neq 0$, with $c_1$ turned on during real time evolution.

\section{Experimental Feasibility}

To estimate the energy and the observability of vortices,
we take $^{87}{\rm Rb}$ atoms as an example (atomic mass $87$u). For a harmonic trapping potential with
typical frequency of $\omega$, $^{87}\text{Rb}$ atoms have
typical length scales $l_{s} = \sqrt{\hbar/M\omega} \simeq 1.1\,(2\pi\times 100\,{\rm Hz}/\omega)^{1/2}\text{$\mu$m}$. With $\omega_{B} = 2g_F\mu_{B}B_1l_{s}/\hbar \simeq 2\times10^3g_F(B_1/{\rm G}\,{\rm cm}^{-1})(2\pi\times 100\,\rm Hz/\omega)^{1/2}Hz$ as the Larmor frequency, strength of the Zeeman coupling is
\begin{align}
\alpha = 2\omega_{B}/\omega \simeq 3\,g_{F}\left(\frac{B_1}{{\rm G}\,{\rm cm}^{-1}}\right)\left(\frac{2\pi\times100\,{\rm Hz}}{\omega}\right)^{3/2}.
\end{align}
The $2$-body interaction scattering lengths are $a_{2}\approx 100\,a_{B}$ and $a_{0} \approx 102\,a_{B}$, where $a_{B} \approx 5.3\times 10^{-2}\,\text{nm}$ is the Bohr radius. This gives $\bar{c}_0 \approx 6.7\times 10^{-2}\,\mu{\rm m}/87{\rm u}$ and $\bar{c}_2 \approx -4.4\times 10^{-4}\,\mu{\rm m}/87{\rm u}$ for the interaction parameters (spin interactions are suppressed by $\simeq 6.5\times 10^{-3}$ as compared to density interactions). Their re-scaled versions are 
\begin{align}
    c_0 \simeq 6.3\times 10^2\left(\frac{N}{10^4}\right)\left(\frac{\omega}{2\pi\times 100\,{\rm Hz}}\right)^{1/2} \;;\qquad c_2 \simeq -4.1\left(\frac{N}{10^4}\right)\left(\frac{\omega}{2\pi\times 100\,{\rm Hz}}\right)^{1/2} 
\end{align}
where $N$ is the total number of atoms in the condensate. The energy scale of the contact interaction per particle on the sphere (of radius $r_0 = l_s\alpha F$ and width $l_s$), is $\mathcal{E}_{c_0} \simeq 0.5\,\bar{c}_0(4\pi r_0^2)l_s (N/(4\pi r_0^2 l_s))^2/N$:
\begin{align}
    \mathcal{E}_{c_0} \simeq \frac{25}{\alpha^2 F^2}\left(\frac{N}{10^4}\right)\left(\frac{\omega}{2\pi\times 100\,{\rm Hz}}\right)^{1/2}\,\hbar\omega\,.
\end{align}
On the other hand, the energy gap per particle, between the lowest ($\tilde{n}=0$) and next ($\tilde{n}=1$) Landau levels is $\Delta E_{\rm LL}\simeq \hbar\omega$. We see that for the chosen parameters $N = 10^4$ and $\omega = 2\pi\times 10$ Hz for $^{87}{\rm Rb}$ atoms ($F = 1$), having $\alpha > 5$ renders $\mathcal{E}_{c_0} < \Delta E_{\rm LL}$. In this case, the effect of interactions can be neglected, and the simpler ``single particle" analysis becomes valid.

\end{document}